\documentclass[aps,prd,nofootinbib,floatfix,twocolumn]{revtex4-2}
\usepackage{amsmath,amssymb,color,graphicx,bm,hyperref,mathrsfs}
\usepackage{physics}
\usepackage{verbatim}

\usepackage{bigints}
\usepackage{float}
\usepackage{booktabs}
\usepackage{graphicx,subfigure,multirow,stackengine}

\definecolor{red}{rgb}{0.8,0,0}
\definecolor{RED}{rgb}{0.8,0,0}
\definecolor{violet}{rgb}{0.4,0,0.4}
\definecolor{green}{rgb}{0,0.5,0.0}
\definecolor{GREEN}{rgb}{0,0.5,0.0}
\definecolor{navy}{rgb}{0.0,0.0,0.6}
\definecolor{orange}{rgb}{0.8,0.2,0.0}
\definecolor{blue}{rgb}{0.3,0.0,0.8}

\begin{document}
\title{\textbf{Influence of thermal bath on Pancharatnam-Berry phase in an accelerated frame}}
\author{Debasish Ghosh}
\email{debasishghosh055@gmail.com}
\author{Bibhas Ranjan Majhi}
\email{bibhas.majhi@iitg.ac.in}
\affiliation{Department of Physics, Indian Institute of Technology Guwahati, Guwahati 781039, Assam, India.}

\begin{abstract}
A uniformly accelerated atom captures Pancharatnam-Berry phase in its quantum state and the phase factor depends on the vacuum fluctuation of the background quantum fields. We observe that the thermal nature of the fields further affects the induced phase. Interestingly the induced phase captures the exchange symmetry between the Unruh and real thermal baths. This observation further supports the claim that the Unruh thermal bath mimics a real thermal bath. Moreover for certain values of system parameters and at high temperature, the phase is enhanced compared to zero temperature situation. However the required temperature to observe the phase experimentally is so high that the detection of Unruh effect through this is not possible within the current technology.
\end{abstract}
\maketitle

\section{Introduction}
A uniformly accelerated observer perceive the Minkowski vacuum as thermal bath with temperature is proportional to acceleration, known as Unruh effect \cite{Fulling:1972md,Davies:1974th,Unruh:1976db}. It appears to be an important theoretical prediction to understand the quantum properties of black hole due to its strong resemblance with the Hawking effect \cite{Hawking:1974rv,Hawking:1975vcx}. Therefore  for better understanding the experimental detection of Unruh effect would be more beneficiary. However the requirement of high acceleration ($a\sim 10^{21}~m/s^2$ for $1$ Kelvin temperature \cite{Crispino:2007eb}) is hindering the success. There are efforts to construct a laboratory apparatus at low acceleration \cite{Unruh:1980cg,Garay:1999sk,Nation:2009xb,Vanzella:2001ec,Scully:2003zz,Lochan:2019osm,Stargen:2021vtg,Chen:1998kp,Schutzhold:2008zza,Retzker:2007vql,Aspachs:2010hh,Lynch:2019hmk,Barman:2024jpc} and the proposed models \cite{Lochan:2019osm,Stargen:2021vtg,Barman:2024jpc} have been able to lower the acceleration as low as $\sim 10^9~m/s^2$. Among these models some of them are based on the detection of induced Pancharatnam-Berry phase (PBP) due to the acceleration of a two-level atom \cite{Martin-Martinez:2010gnz,Hu:2012ed}. This indirect way of detecting Unruh effect appears to be more convenient as it requires reasonably less acceleration for a specific arrangement \cite{Barman:2024jpc}. Therefore deeper understanding of PBP, induced in the qubit due to its acceleration, is very important not only for the experimental purpose but also for theoretical point of view.

PBP is a geometrical phase, first introduced by Pancharatnam \cite{Pancharatnam:1956url} in his study of the interference of classical light in a distinct state of polarisation. Some time later Berry \cite{Berry:1984jv} introduced the phase which is a quantum counterpart of the Pancharatnam phase in the case of cyclic adiabatic evolution. Using open quantum system formalism, PBP has been calculated first for an accelerated two-level atom with an electric dipole, interacting with background electric field \cite{Hu:2012ed}. It appeared that the two point correlation of the electric field induces PBP to the atom initial state. However the measurable minimum phase has appeared to be achieved with acceleration $a\sim 10^{18} ~m/s^2$ for the existing experimental precision. Further the same model has been studied various situations \cite{Hu:2012gv,Tian:2013lna,Jin:2014spa,Cai:2018mbp,Wang:2019hlr,Quach:2021vzo,Zhao:2022zbu,Arya:2022lay}, like motion of atom on dS spacetime \cite{Tian:2013lna,Tian:2014jha}, circular motion on Minkowski spacetime \cite{Jin:2014spa,Arya:2022lay}, uniform accelerated motion with one mirror and double mirrors \cite{Barman:2024jpc}, etc. However, all these studies considered the background fields are non-thermal.

From practical point of view, the environment is always at finite temperature and therefore the background fields must be within in a ``real'' thermal bath. In various situations, it has been observed that the bath temperature has significant influence on the transition rate of the detector \cite{Costa:1994yx,Hodgkinson:2014iua,Kolekar:2013xua,Kolekar:2013hra,Chowdhury:2019set,Barman:2021oum,Barman:2022utm} and on the entanglement phenomenon between two such qubits \cite{Barman:2021bbw,Chowdhury:2021ieg}. Since the PBP is mostly dictated by the detector's transition rate, we expect to see the thermal nature of field will influence the induced PBP of the qubit's quantum state. Hence it would be interesting to check how such temperature can be a factor in determining the required acceleration of the detector to achieve the minimum measurable phase.

Apart from this experimental importance, there is theoretical importance as well. Due to the background temperature, the transition rate of the detector is contributed not only by the {\it induced transition} due to its acceleration, but also by the {\it spontaneous transition} influenced by the temperature. Moreover, the temperature also influences another transition on the accelerated transition. Therefore the full transition rate is sum of two induced transitions and one spontaneous transition \cite{Kolekar:2013aka,Kolekar:2013hra,Kolekar:2013xua}. Moreover, this is symmetric under the interchange of Unruh temperature and that of the background thermal bath. Such an observation implies that the Unruh thermal bath can play the role of a real thermal bath. We like to investigate how the PBP phase will be influenced and whether such properties are also apparent in this case as well. 

Following these two fold aims, we calculate the PBP of a uniformly accelerated (with acceleration $a$) qubit which is interacting with background real scalar field at inverse temperature $\beta$. We observe that both acceleration and temperature induced PBP, like the transition amplitude, is decomposed into three parts: one is due to pure background temperature, second part depends purely on acceleration and other one is influenced by both acceleration and temperature. Interestingly the induced PBP follows the same properties like the detector's transition rate. This observation has a theoretical importance -- it shows, as far as PBP is concerned, that the Unruh thermal bath and real thermal bath can be put in equal footing and thereby bolstering the earlier claim, obtained through the transition rate.  

By subtracting the pure thermal contribution in the PBP, the thermal-induced accelerated part is being obtained. We observe that for certain fixed parameter values (like acceleration and energy gap between the two levels of qubit), the thermal-induced accelerated PBP can be larger than pure thermal or pure accelerated PBPs. For large temperatures, this is much prominent. We further show that although such enhancement is possible, but it is not enough for experimental detection within the current technology. 

\section{The PBP for a two-level open quantum system: a brief overview}
Here we present a schematic idea to calculate the PBP, induced in a two-level quantum system, when it is interacting with a real scalar field. Although it is known in literature \cite{Hu:2012ed}, but for few clarifications and consistency of its applicability we mention the main features so that we can give our argument on the choice of correct correlation function when the field is in real thermal bath.  
For the adiabatic change with time in the Hamiltonian of a quantum system, the state of the system changes by a phase factor. 
If the time-dependent parameters of the Hamiltonian make a closed loop in the parameter space under the adiabatic approximation then the phase acquired by the system depends on the geometry of this parameter space. After the introduction by Pancharatnam, Berry followed this, and therefore in literature, it is known as the PB phase.
For a quantum system, this is given by the following general expression \cite{Pancharatnam:1956url,Berry:1984jv}
 \begin{equation}
 \label{eq11}
  \Phi(t) = i \int_{0}^{t} \braket{\psi_k(t\sp{\prime})}{\dot{\psi}_k(t\sp{\prime})} dt\sp{\prime}~.
 \end{equation}
In the above $\ket{\psi_k (t)}$ is the $k^{th}$ eigenstate of the system at time $t$ and $\ket{\dot{\psi}_k(t)} \equiv \frac{\partial}{\partial t} \ket{\psi_k(t)}$. Here we intend to calculate $\Phi(t)$ for a two-level quantum system which is interacting with an environment. Therefore, below we will briefly introduce the main necessary ingredients related to the open quantum system.

Denote the system ($S$) and the environment ($B$) Hamiltonians as $H_S$ and $H_B$, respectively. Then the total Hamiltonian of the system-environment composite system is given by $H(t) = H_s \otimes I_B + I_s \otimes H_B + H_I(t)$.
Here $H_I(t)$ denotes the interaction Hamiltonian, which consists of both the system and environment operators. 
In our model, the system is a two-level atom whose Hamiltonian is given by $H_s = \dfrac{1}{2} \hbar \omega_0 \sigma_3$, where $\sigma_3$ is third component of the Pauli matrices and $\omega_0$ is the energy gap between two levels. The environment is considered to be quantum real scalar fields $ \phi(x)$ and the interaction between $S$ and $B$ is taken to be $H_I(t) = g \phi(x) \sigma_2 $, where $\sigma_2$ is the second component of Pauli metrices. We consider initially the field was in vacuum state while the system was in state denoted by $\ket{\psi(0)}$. 
 Assuming the interaction between $S$ and $B$ is fragile; i.e. in the weak coupling limit, the evolution of the reduced density matrix of $S$ can be determined from the Kossakowski-Lindblad equation (in Schrodinger picture) \cite{Hu:2012ed} 
\begin{equation} \label{eq23}
\dfrac{d\rho}{dt} = -\dfrac{i}{\hbar} [H_{\text{eff}}(t),\rho(t)] + D[\rho(t)]~.
\end{equation}
$H_{\text{eff}} = H_s+H_{LS}$ is the sum of the system Hamiltonian and Lamb shift Hamiltonian. 
$H_{LS}$ leads to a renormalization of the system's unperturbed energy levels induced by the system-environment interaction and $D[\rho(t)]$ is the dissipation. 
 
 In this case the Lamb shift Hamiltonian is, $H_{LS} = \dfrac{1}{2} \hbar \Delta \sigma_3$, where $\Delta$ is Lamb shift factor. Hence $H_{\text{eff}}$ is given by $H_{\text{eff}} = \dfrac{1}{2} \hbar \Omega \sigma_3$, where $\Omega = \omega_0+\Delta$. Usually $\Delta<<\omega_0$ and hence we will neglect $\Delta$ in the latter analysis.
For a two level system, dissipation term takes the form
\begin{equation}
\label{eq27}
D[\rho(t)] = \dfrac{1}{2} \sum_{i,j = 1}^{3} a_{ij} [ 2 \sigma_j \rho \sigma_i - \sigma_i \sigma_j \rho - \rho \sigma_i \sigma_j]~,
\end{equation}
where $a_{ij}$ is called as the Kossakowski matrix. According to our system Hamiltonian, it has a form like,
\begin{equation}\label{k1}
    a_{ij} = A \delta_{ij} - iB \epsilon_{ijk} \delta_{k3} + C\delta_{i3} \delta_{j3}~.
\end{equation}
Here $a_{ij}$ is function of $\omega_0$. The coefficients $A$ and $B$ are
\begin{eqnarray}
\label{eq29}
&&A = \dfrac{g^2}{2} [ \gamma(\omega_0) + \gamma(-\omega_0)],  B = \dfrac{g^2}{2} [\gamma(\omega_0) - \gamma(-\omega_0)]~.
\end{eqnarray}
On the other hand for our choice of the interaction Hamiltonian ($H_I(t) = g \phi(x) \sigma_2 $), one finds $C=-A$. A more general discussion was given much earlier in \cite{PhysRevA.70.012112,PhysRevD.77.024031,10.1093/acprof:oso/9780199213900.001.0001}.

In this case $\gamma(\omega)$ boils down to 
\begin{equation}
\label{eq30}
\gamma(\omega) = \int_{-\infty}^{\infty} ds e^{i \omega s} G^+(x,x')~, 
\end{equation}
where $s= (t'-t)$ and $G^+(x,x')$ is known as the positive frequency Wightman function: 
$G^+(x,x') = \bra{0} \phi(t,x) \phi(t\sp{\prime}, x \sp{\prime}) \ket{0}$.
These results are derived using three approximations, namely Born, Markov and rotating-wave approximations.
A detailed discussion can be followed from \cite{10.1093/acprof:oso/9780199213900.001.0001}. Note that the two-point correlation function in (\ref{eq30}) is time translational invariant. This is due the fact that the interaction has been considered as Markov process where the interaction time is much larger than the inherent time scale of the system.  
Therefore $G^+(x,x')$ is function of $(t'-t)$ and so it should not depend on the initial and final times.

Consider the initial state of the system $S$ as
\begin{equation}
 \ket{\psi(0)} = \cos\dfrac{\theta}{2} \ket{0} + \sin\dfrac{\theta}{2} \ket{1}~.
\end{equation}
Here $\ket{0}$ and $\ket{1}$ denote ground and excited states of $S$.
\begin{widetext}
Then solving (\ref{eq23}) one finds the density matrix as \cite{Hu:2012ed} 
\begin{equation}
    {\rho(\tau)} =
    \begin{bmatrix}
        \exp{-4A\tau} \cos^2\dfrac{\theta}{2} + \dfrac{(B-A)}{2A}(\exp{-4A\tau} -1) & \dfrac{1}{2} \sin\dfrac{\theta}{2} \exp{-2A\tau - i\Omega \tau} \\
        \\
        \dfrac{1}{2} \sin\dfrac{\theta}{2} \exp{-2A\tau + i\Omega \tau} &  1 -\exp{-4A\tau} \cos^2\dfrac{\theta}{2} - \dfrac{(B-A)}{2A}(\exp{-4A\tau} -1)
    \end{bmatrix}~.
\end{equation}
Here we use time $t$ as the atom's proper time $\tau$.
The eigenvalues of the density matrix are $\lambda = \dfrac{1}{8} (4 \pm \sqrt{\chi})$
where 
\begin{equation}
    \chi = 16  \exp{-4A\tau} \sin^2{\theta} + 16 \left[ (\dfrac{B}{A}) (\exp{-4A\tau} -1) + \exp{-4A\tau} \cos\theta \right]^2~.
\end{equation}
The corresponding eigenvectors are, 
\begin{eqnarray}
&&\ket {\psi_1 (\tau)} = \sin\dfrac{\theta_{\tau}}{2} \ket{0} + \exp{i\Omega \tau} \cos\dfrac{\theta_{\tau}}{2} \ket{1}~;
\nonumber
\\
&&\ket {\psi_2 (\tau)} = \cos\dfrac{\theta_{\tau}}{2} \ket{0} + \exp{i\Omega \tau} \sin\dfrac{\theta_{\tau}}{2} \ket{1}~,
\label{B1}
\end{eqnarray}
where 
\begin{equation}
\tan\theta_{\tau} = \sqrt{\dfrac{\eta_0 + P}{\eta_0 - P}}~; \,\,\,\ P =  \left[ \Big(\dfrac{B}{A}\Big) \Big(\exp{-4A\tau} -1\Big) + \exp{-4A\tau} \cos\theta \right]~; \,\,\,\,\,\ \eta_0 = \sqrt{P^2 + \exp{-4A\tau} \sin^2{\theta}}~.
\end{equation}
\end{widetext}

Now we can calculate the PBP by integrating equation (\ref{eq11}) over a full cycle; i.e. choose the upper limit as $ \dfrac{2 \pi}{\omega_0}$, where the states are given by (\ref{B1}). So the general expression for phase in terms of $A$ and $B$ is
\begin{eqnarray}
    \Phi = -\pi(1-\cos\theta) - \dfrac{2 \pi^2 B}{\omega_0} \sin^2\theta \left[ 2+ \dfrac{A}{B} \cos\theta \right]~.
    \label{T2}
\end{eqnarray}
Now our task is to calculate the coefficients $A$ and $B$ (given by (\ref{eq29})) to obtain the phases. Note that any one of the final states (\ref{B1}) is enough to obtain the phase as other one will give the same result. We considered the first one.

\section{Berry-Pancharatnam phase due to thermal field}
In this section, we want to derive PB phase acquired by an accelerated observer which is in a thermal bath. Therefore we first find the form of the thermal Wightman function.

\subsection{Thermal Wightman Function with Minkowski modes}
We take our system to be in equilibrium with a thermal bath characterized by the parameter $\beta = \dfrac{1}{T}$, where $T$ the temperature of the thermal bath. Then the thermal Wightman function is defined as, $G^+_{\beta} (x_2,x_1) = \dfrac{1}{Z} \Tr{ \exp{-\beta H} \phi(x_2) \phi(x_1)}$,
where, $Z = \Tr{ \exp{-\beta H}}$.
In Fourier domain the Hamiltonian of scalar fields can be cast as a sum of infinitely many simple harmonic oscillator and so, $H = \sum
_k a_k^\dagger a_k \omega_k$. Using these and the mode expansion of the scalar field with respect to Minkowski modes in $(3+1)$-spacetime dimensions, one finds the Wightman function as \cite{Weldon:2000pe,Chowdhury:2019set}
\begin{equation}
\label{eq54}
G^+_{\beta} (x_2,x_1) = \int \dfrac{d^3k}{(2\pi)^3 2 \omega_k} \left[ \dfrac{e^{i \vec{k}\cdot \Delta \Vec{x} + i \omega_k \Delta t }}{e^{\beta \omega_k}-1} - \dfrac{e^{i \vec{k}\cdot \Delta \Vec{x} - i \omega_k \Delta t }}{e^{-\beta \omega_k}-1}\right]~.
\end{equation}
In the above we defined $\Delta \Vec{x} = \vec{x}_2- \vec{x}_1$ and $\Delta t = t_2 -t_1$. Since our system $S$ is accelerating we need to transform these Minkowski coordinates in terms of Rindler proper time. The relations between Minkowski coordinates $(t,\vec{x})$ and Rindler coordinates $(\eta,\xi)$, for the system moving along $x$-axis, is given by 
\begin{equation}
t= \dfrac{e^{a\xi}}{a} \sinh{a\eta} ,   x = \dfrac{e^{a\xi}}{a} \cosh{a\eta}~.
\label{B2}
\end{equation}
Note that substitution of these in (\ref{eq54}) with $\xi=0$ (as Rindler frame is the proper frame) does not keep the thermal Wightman function time translational invariant with respect to Rindler proper time (same was also noticed earlier in \cite{Chowdhury:2019set,Barman:2021oum}). Therefore the formalism introduced in previous section is not applicable for the analysis with Minkowski mode decomposition of the scalar field. We will see in next that Unruh mode decomposition is more suitable to discuss the background thermal effect on PB phase. Similar has also been noticed in various investigation (see e.g. \cite{Kolekar:2013aka,Barman:2021oum,Das:2022qcr}).

\begin{widetext}
\subsection{Wightman function with Unruh modes}
For the system accelerating on the right Rindler wedge (RRW), using the decomposition of the field in terms of Unruh modes \cite{book:Birrell}
%is given by
%\begin{eqnarray}
%\phi_R(X) &=& \sum_{\omega=0}^{\infty}\sum_{k=-\infty}^{\infty} \dfrac{1}{\sqrt{2\sinh{\dfrac{\pi \omega}{a}}}} \Big[ d^1_{\omega, k_p} e^{\dfrac{\pi \omega}{2a}} {^R {u}_{\omega,k_p}} 
%\nonumber
%\\ 
%&+& d^2_{\omega, k_p} e^{-\dfrac{\pi \omega}{2a}} {^R {u}^*_{\omega,-k_p}} \Big] + h.c~.
%\end{eqnarray}
%Here the operator $\left(d^1_{\omega, k_p},d^{1 \dagger}_{\omega, k_p}\right) $ and $(d^2_{\omega, k_p},d^{2\dagger}_{\omega, k_p})$ are the two sets of annihilation and creation operators for the Unruh modes. The annihilation operators annihilates the Minkowski vacuum: $d^1_{\omega, k_p} \ket{0_M} = d^2_{\omega, k_p} \ket{0_M} = 0$.
%For the system accelerating on left Rindler wedge (LRW), the composition of field in terms of Unruh modes is given by 
%\begin{eqnarray}
%\phi_L(X) &=& \sum_{\omega=0}^{\infty}\sum_{k=-\infty}^{\infty} \dfrac{1}{\sqrt{2\sinh{\dfrac{\pi \omega}{a}}}} \Big[ d^1_{\omega, k_p} e^{-\dfrac{\pi \omega}{2a}} {^L {u}^*_{\omega,-k_p}} 
%\nonumber
%\\ 
%&+& d^2_{\omega, k_p} e^{\dfrac{\pi \omega}{2a}} {^L {u}_{\omega,k_p}} \Big] + h.c~,
%\end{eqnarray}
%where the operator $\left(d^1_{\omega, k_p},d^{1 \dagger}_{\omega, k_p}\right) $ and $(d^2_{\omega, k_p},d^{2\dagger}_{\omega, k_p})$ are same as before (see \cite{book:Birrell,Barman:2021oum,Barman:2021bbw} for details). \textcolor{red}{The mode functions are given by....}
one finds the positive frequency thermal Wightman function on the Rindler proper frame as
\cite{Barman:2021oum,Barman:2021bbw}
\begin{equation}
G^+_{\beta R}  = \int_0^{\infty} d\omega \int \dfrac{d^2k_p}{(2 \pi)^4} \dfrac{2}{a} \Big( \dfrac{e^{-i \omega \Delta\eta}  e^{\dfrac{\pi \omega}{a}}+ e^{i \omega \Delta\eta}   e^{-\dfrac{\pi \omega}{a}}}{1-e^{-\beta \omega}}
 + \dfrac{e^{i \omega \Delta\eta}  e^{\dfrac{\pi \omega}{a}}+ e^{-i \omega \Delta\eta}   e^{-\dfrac{\pi \omega}{a}}}{e^{\beta \omega}-1} \Big)  \kappa\Big[ \dfrac{i\omega}{a}, \dfrac{k_p e^{a\xi}}{a}\Big] \kappa\Big[ \dfrac{i\omega}{a}, \dfrac{k_p e^{a\xi}}{a}\Big]~.
 \label{B3}
\end{equation}
Here $\kappa[ n , z]$ denotes the modified Bessel function of the second kind of order $n$. %In the above we have chosen $\Delta \vec{x}_p = 0$ and also $\xi$ has been taken to be same (which will be put to zero in the later analysis) as will be working in proper frame. 
Here $\eta$ is identified as proper time $\tau$.
Note that the above one is manifestly time-translational invariant with respect to Rindler proper time.
While the thermal Wightman function in LRW case is given by \cite{Barman:2021oum,Barman:2021bbw},
\begin{equation}
G^+_{\beta L}  = \int_0^{\infty} d\omega \int \dfrac{d^2k_p}{(2 \pi)^4} \dfrac{2}{a} \Big( \dfrac{e^{i \omega \Delta\eta}  e^{\dfrac{\pi \omega}{a}}+ e^{-i \omega \Delta\eta}   e^{-\dfrac{\pi \omega}{a}}}{1-e^{-\beta \omega}}
 + \dfrac{e^{-i \omega \Delta\eta}  e^{\dfrac{\pi \omega}{a}}+ e^{i \omega \Delta\eta}   e^{-\dfrac{\pi \omega}{a}}}{e^{\beta \omega}-1} \Big)  \kappa\Big[ \dfrac{i\omega}{a}, \dfrac{k_p e^{a\xi}}{a}\Big] \kappa\Big[ \dfrac{i\omega}{a}, \dfrac{k_p e^{a\xi}}{a}\Big]~.
\end{equation}

\end{widetext}

\subsection{PB phase}
The $A$ and $B$ coefficients can be calculated using Eqs. (\ref{eq29}) and (\ref{eq30}). Substituting (\ref{B3}) into (\ref{eq30}) one finds (see Appendix \ref{App1}),
\begin{equation}
    \gamma(\omega_0) = \dfrac{\omega_0}{2\pi} \dfrac{\left[ \dfrac{e^{\dfrac{\pi \omega_0}{a}}}{1 - e^{-\beta \omega_0}} +  \dfrac{e^{-\dfrac{\pi \omega_0}{a}}}{e^{\beta \omega_0}-1}\right]}{2 \sinh{\dfrac{\pi \omega_0}{a}}}~;
    \label{T3}
\end{equation}
and
\begin{equation}
    \gamma(-\omega_0) = \dfrac{\omega_0}{2\pi} \dfrac{\left[ \dfrac{e^{-\dfrac{\pi \omega_0}{a}}}{1 - e^{-\beta \omega_0}} +  \dfrac{e^{\dfrac{\pi \omega_0}{a}}}{e^{\beta \omega_0}-1}\right]}{2 \sinh{\dfrac{\pi \omega_0}{a}}}~.
    \label{T31}
\end{equation}
Therefore we find
\begin{eqnarray}
&&A = \chi_0 \coth{\dfrac{\omega_0 \pi}{a}} \coth{\dfrac{\beta \omega_0}{2}}~;
\nonumber
\\
&&B = \chi_0~, 
\end{eqnarray}
where $\chi_0 = \dfrac{ g^2 \omega_0}{4 \pi}$.
So the PBP acquired by an accelerated observer in thermal bath comes out to be (using the above expressions in (\ref{T2}))
\begin{eqnarray}
\Phi &=& -\pi (1-\cos{\theta}) 
\nonumber
\\
&-& 2 \pi^2 \dfrac{\chi_0}{\omega_0} \sin^2\theta \Big[ 2+ \coth{\dfrac{\omega_0 \beta_U}{2}} \coth{\dfrac{\beta \omega_0}{2}} \cos{\theta}\Big]~,
\end{eqnarray}
where $\beta_U = \dfrac{1}{T_U} = \dfrac{2 \pi}{a}$ is the inverse Unruh temperature \cite{Unruh:1976db}. 
Note that for both $\beta\to\infty$ and $\beta_U\to\infty$, there is a non-vanishing contribution to phase
\begin{equation}
\Phi_{(0,0)} = -\pi (1-\cos{\theta}) - 2 \pi^2 \dfrac{\chi_0}{\omega_0} \sin^2\theta \Big[2+ \cos{\theta}\Big]~.
\label{B4}
\end{equation}
Therefore the only contribution from simultaneous effects of acceleration and background thermal bath is given by
\begin{eqnarray}
&&\delta\Phi = \Phi - \Phi_{(0,0)} 
\nonumber
\\
&=&  - 2 \pi^2 \dfrac{\chi_0}{\omega_0} \sin^2\theta \Big[\coth{\dfrac{\omega_0 \beta_U}{2}} \coth{\dfrac{\beta \omega_0}{2}} - 1\Big] \cos{\theta}
\nonumber
\\
&=&  - \pi g^2 \sin^2\theta \Big[n_\beta + n_{\beta_U} 
+ 2 n_\beta n_{\beta_U}\Big] \cos{\theta}~,
\label{B5}
\end{eqnarray}
where
\begin{equation}
n_\beta = \frac{1}{e^{\beta\omega_0} - 1}~; \,\,\,\,\
n_{\beta_U} = \frac{1}{e^{\beta_U\omega_0} - 1}~.
\end{equation}

Note that (\ref{B5}) is symmetric under $\beta\leftrightarrow \beta_U$. Moreover it vanishes for two situations: (i) $\theta = 0$ and (ii) $\theta = \pi/2$. The conditions, (i) and (ii), denote either the initial state is ground state or the excited state, respectively. It is interesting to observe that the phase is contributed not only purely from thermal bath and motion of the system, there is induced one as well. The last term of the above is the induced phase in the accelerated system due to the presence of thermal bath or vice-versa. Therefore we call the first two terms as the {\it spontaneous} and induced (due to acceleration only) phase factors while the last one is called as the {\it stimulated} phase factor. Similar situation was also observed in the transition probability of a Unruh-DeWitt detector which is interacting with the thermal fields \cite{Kolekar:2013xua,Kolekar:2013hra}. This particular observation has an important theoretical importance. It further solidified the idea of equivalence between the real thermal bath and Unruh thermal bath which goes beyond the transition rate of a Unruh-DeWitt detector \cite{Kolekar:2013xua} or particle production in Minkowski vacuum \cite{Kolekar:2013hra}. 

To understand the enhancement due to the background thermal bath, let us now subtract the sole contribution of pure thermal effect. In this case the phase is contributed from the following term:
\begin{equation}
\delta\Phi_{\beta_U} =   -  \pi g^2 \sin^2\theta \Big[n_{\beta_U} 
+ 2 n_\beta n_{\beta_U}\Big] \cos{\theta}~.
\label{B6}
\end{equation} 
This can be interpreted as the {\it Unruh-thermal induced effect} in the phase. The relative contribution to the Unruh-thermal induced effect with respect to pure thermal effect is then
\begin{equation}
\delta\Phi_{\text{rel}\beta} = \frac{e^{\beta\omega_0} + 1}{e^{\beta_U\omega_0} - 1}~.
\label{B7}
\end{equation} 
On the other hand if we investigate how much enhancement in phase due to Unruh-thermal induced effect has been achieved with respect to that with only accelerated motion, then one finds
\begin{equation}
\delta\Phi_{\text{relU}} = \frac{\delta\Phi_{\beta_U}}{\delta\Phi^0_{\beta_U}} = \coth\Big(\frac{\beta\omega_0}{2}\Big)~,
\label{B8}
\end{equation} 
where ${\delta\Phi^0_{\beta_U}}  = -  \pi  n_{\beta_U}\sin^2\theta  \cos{\theta}$.
Surprisingly, the above is independent of the value for the acceleration of the system. This implies that the phase of the system's quantum state, affected by Unruh-thermal induced effect, will enhance for its any non-vanishing accelerated motion by a factor which depends only on the background temperature compared to only Unruh effect (i.e. zero temperature background case). 
Now since for $\frac{\beta\omega_0}{2}\geq 0$ we have $1\leq \coth(\frac{\beta\omega_0}{2})< \infty$, at low temperature the phase will be small ($\coth(\frac{\beta\omega_0}{2})\to 1$ as $\beta\to\infty$); while at high temperature the phase will be large ($\coth(\frac{\beta\omega_0}{2})\to\infty$ as $\beta\to 0$). Hence the phase due to Unruh-thermal induced effect (subtracting the pure thermal part) will enhance as the background temperature increases (see Eq. (\ref{B6})).

\begin{figure}[htbp!]
\centering
\includegraphics[width=1.0\linewidth]{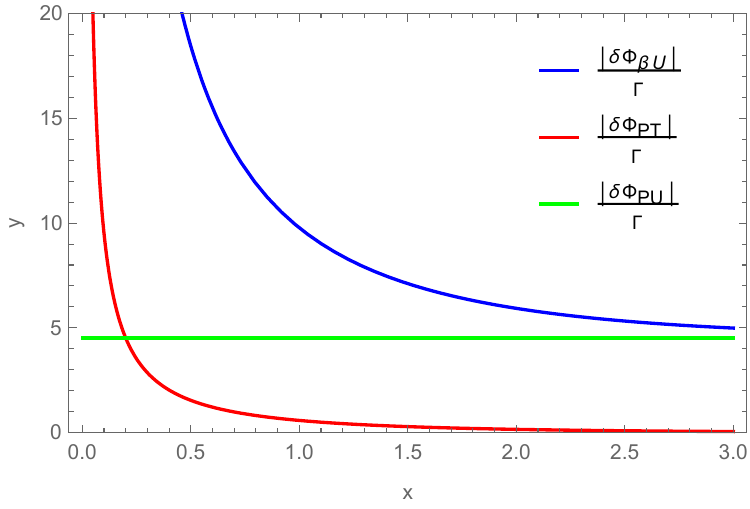}
\caption{Variation of phases for different situations as a function of temperature of the background bath (for $\beta_{U}\omega_0 = 0.2$).}
\label{Plot 1:Test123}
\end{figure}

\begin{figure}[htbp!]
\centering
\includegraphics[width=1.0\linewidth]{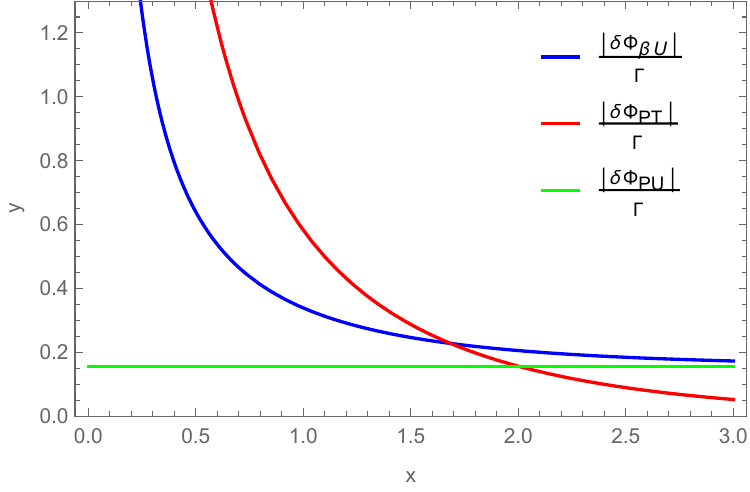}
\caption{Variation of phases for different situations as a function of temperature of the background bath (for $\beta_{U}\omega_0 = 2$).}
\label{Plot 2:Test123}
\end{figure}

We now plot absolute quantities, Unruh-thermal induced
($\delta\Phi_{\beta_U}$), pure thermal ($\delta\Phi_{PT}$) and pure Unruh  ($\delta\Phi_{PU}$) phases (along $y$-axis) per unit $\Gamma ( =  \dfrac{\pi g^2}{2\sqrt{2}} $), as a function of $x=\omega_0\beta$ for $\theta=\pi/4$. The quantities are given by
\begin{eqnarray}\label{eq37}
&&\delta\Phi_{\beta_U} =   - \dfrac{\pi g^2}{2\sqrt{2}}\Big[\frac{1}{e^{\beta_U\omega_0} - 1} 
+ \frac{2}{(e^{\beta_U\omega_0} - 1) (e^{\beta\omega_0} - 1)}\Big]~;
\\
&&\delta\Phi_{PT} = - \dfrac{\pi g^2}{2\sqrt{2}} \Big(\frac{1}{e^{\beta\omega_0} - 1}\Big
)~;
\\
&&\delta\Phi_{PU} = - \dfrac{\pi g^2}{2\sqrt{2}} \Big(\frac{1}{e^{\beta_U\omega_0} - 1}\Big)~.
\end{eqnarray}
Fig \ref{Plot 1:Test123} corresponds to $\beta_{U}\omega_0 = 0.2$ and Fig. \ref{Plot 2:Test123} is drawn with $\beta_{U}\omega_0 = 2$. 
%%%%%%%%%%%%%%%%%%%%%%%%%%%%%%%%%%%%%%%%%%%%%%%%%%%%%
We found that for a particular parameters the phase is enhanced by the background temperature, specifically when the acceleration is low but the temperature is very high (see Fig. \ref{Plot 1:Test123}). However for other parameter choices, the situation may change (e.g. see Fig. \ref{Plot 2:Test123}).

Similarly as previous case we can find PBP of the qubit when it is moving in the left-Rindler wedge (LRW). In this case $\gamma(\omega_0)$ is given by that for RRW (see Eq. (\ref{T3})). 
Therefore the acquired phases in the qubit's quantum state are same for both the motions.

Note that the results are obtained by using the Unruh mode decomposition. Ideally one should use the Minkowski mode decomposition. But we know that each Minkowski operator (creation or annihilation) is function of both the Rindler operators. Therefore a Minkowski mode with positive frequency is a mixture of both positive and negative frequencies Rindler modes. This results in non-translational invariance with respect to Rindler time in Wightman function when one uses Minkowski mode decomposition. Thus non-stationary situation arises. Such a difficulty can be avoided by using Unruh modes, which are linear combination of Rindler modes in such a way that here positive and negative frequencies do not get mixed. Therefore the vacuum state of Minkowski mode (namely Minkowski vacuum) remains same for the Unruh mode. In this sense, as far as Minkowski vacuum is concerned, the Unruh mode is reminiscent to Minkowski one and can be used as a proxy for Minkowski counter parts. Moreover this avoids the issue of non-stationarity and hence fits well within our formalism. However this does not fulfill the perspective of the inertial observer. For the present analysis the particle production is described as the excitation of Unruh mode which is contrary to the usual notion of excitation of Minkowski mode. But as both these modes share the same vacuum, the present result can be attributed as the effect of Minkowski vacuum fluctuation. In summary, although the Unruh particles are different from the Minkowski ones, both belong to excited states of the same vacuum. Then as we are interested on the thermal state (as the background must be thermal in reality), the choice of Unruh mode decomposition is equally well for the present purpose.  

\section{Experimental feasibility: comparison with non-thermal background}
We now compare the required acceleration between the thermal and non-thermal backgrounds for a minimal measurable PBP of the order $\sim 10^{-5}$ \cite{Wang:18} and want to see whether the background thermal effect can help to reach the acceleration at an experimentally possible value.
The PBP for non-thermal background (including all $\hbar$ and c) is,
\begin{eqnarray}
    \delta \Phi_{PU} = -\dfrac{\pi g^2}{2 \hbar c^3} \sin^2\theta \cos\theta \Big[ \coth\dfrac{\omega_0 \pi c }{a_{\text{NTh.}}} - 1\Big]~.
\end{eqnarray}
This is obtained by taking $\beta\to\infty$ in (\ref{B5}) and denoting $a(\beta\to\infty) \equiv a_{\text{NTh.}}$.
Then the required acceleration is given by
\begin{eqnarray}
    a_{\text{NTh.}} = \dfrac{2 \pi \omega_0 c}{\ln{\Big[1+\dfrac{\pi g^2 \sin^2\theta \cos\theta }{2 \hbar c^3 |\delta \Phi_{PU}|}}\Big]}~.
\end{eqnarray}
In the similar way derive the acceleration (from (\ref{B5}) ) for the thermal background as 
\begin{eqnarray}
     a_{\text{Th.}} = \dfrac{2 \pi \omega_0 c}{\ln{\Big[1+\dfrac{\pi g^2 \sin^2\theta \cos\theta }{  2 \hbar c^3 |\delta \Phi_{\beta_U}|}\Big( 1+ \dfrac{2}{e^{\beta \omega_0 -1}}\Big) }\Big]}~.
\end{eqnarray}

Now for a flux qubit, we know $\omega_0 \sim 1$ GHz \cite{PhysRevLett.105.060503} and so $H_s \simeq \hbar\omega_0 \sim 10^{-25}$J.
Since the perturbative calculation is valid for $H_{I} << H_s$, we must have $H_{I} << 10^{-25}$J.
So we choose $g \phi << 10^{-25}$J, and in that case we should assign $ g^2 \langle \phi \phi \rangle << 10^{-50}$. We know that $ \langle \phi \phi \rangle \sim \hbar / c \sim 10^{-42}$ and that means  $ g^2 / (\hbar c^3) << 10^{2}$. So we can consider $ g^2/(\hbar c^3) \simeq 1$.
With this for the non-thermal case, to achieve the minimum measurable phase ($\sim 10^{-5}$) with $\theta = \pi/4$, the required acceleration is $ a_{\text{NTh.}} \sim 10^{17}~m/s^2$ .

On the other hand for the thermal case, the additional factor is $ 2/ ( e^{\hbar \omega_0}/{k_B T}-1)$, where $T$ is the temperature of the background field. Considering $T\sim 1000 K$ (e.g. the temperature of sun) one has $2/ ( e^{\hbar \omega_0}/{k_B T}-1)\sim 10^{5}$ and then the required acceleration for minimum measurable phase of flux qubit is $a_{\text{Th.}} \sim 10^{16}~m/s^2$. It shows that the present available temperature is not sufficient to provide a reasonable value of acceleration and therefore at least with the current technology the experimental evidence is far reachable.

\section{Conclusions}
We investigated the effects of background thermal bath on the induced PBP of a qubit which is uniformly accelerating. Calculation showed that for certain choices of parameters of the system, the phase can be enhanced by the background thermal bath. However the bath temperature must be very high. Consequently we found that a reasonable background temperature is not sufficient to find the minimum measurable phase with low acceleration of the qubit. Rather we need a very large acceleration which is not possible to achieve with the current technology.  

Although this analysis does not provide a feasibility to detect Unruh effect through PBP in an experimental setup, but contains striking theoretical importance. A general consensus is -- Unruh thermal bath can mimic a real thermal bath. In various occasions the same has been tested -- by calculating number operator on a Rindler-Rindler frame and comparing with an accelerated Detector response function on a real thermal bath \cite{Kolekar:2013xua,Kolekar:2013hra} and also calculating the density operator for the thermal fields viewed from an accelerated frame \cite{Kolekar:2013aka}. All these results show a very common feature -- the obtained quantities are symmetric under exchange of temperature of Unruh thermal bath and that for the real thermal bath. However this analogy breaks down in certain scenarios, like for the circular motion \cite{Chowdhury:2019set}. The robustness of this claim therefore has to be tested in various occasions. 

The present scenario provides one of such ambiences. We observed that the induced phase takes a structure which is identical to the response function of a Unruh-DeWitt detector, interacting with thermal fields. Therefore the exchange symmetry also visible in PBP and hence further provides an evidence of equivalence at the quantum level. 

The whole analysis has been done based on a Markov process. Therefore we forced to choose two point correlation function related to Unruh decomposition to maintain the time-translation invariance. However it would be interesting to investigate the same within non-Markovian scenario where the Minkowski mode decomposition can be taken into account. 
%\newpage

\begin{widetext}
\appendix
\section{Calculation of $\gamma(\omega_0)$ and $\gamma(-\omega_0)$}\label{App1}
The explicit value of $\gamma(\omega_0)$ can be obtained using (\ref{B3}) in (\ref{eq30}). This, with the choice $s=\Delta\eta$, yields
\begin{eqnarray}
\gamma(\omega_0) & =&\int_{-\infty}^{\infty} d\Delta\eta  e^{i \omega_0  \Delta\eta}  G^+_{\beta R}
\nonumber
\\
         & =& \dfrac{2}{a} \int_{-\infty}^{\infty} d\Delta\eta \int_0^{\infty} d\omega  e^{i \omega_0  \Delta\eta} \left( \dfrac{e^{-i \omega \Delta\eta}  e^{\dfrac{\pi \omega}{a}}+ e^{i \omega \Delta\eta}   e^{-\dfrac{\pi \omega}{a}}}{1-e^{-\beta \omega}}
         +  \dfrac{e^{i \omega \Delta\eta}  e^{\dfrac{\pi \omega}{a}}+ e^{-i \omega \Delta\eta}   e^{-\dfrac{\pi \omega}{a}}}{e^{\beta \omega}-1} \right) 
         \nonumber
         \\
         &\times& \int_0^{\infty} \dfrac{2 \pi k_p dk_p}{(2 \pi)^4} \kappa\left[ \dfrac{i\omega}{a}, \dfrac{k_p e^{a\xi}}{a}\right] \kappa\left[ \dfrac{i\omega}{a}, \dfrac{k_p e^{a\xi}}{a}\right] ~.
    \end{eqnarray}
Now use of the following result
\begin{equation}
   \int_0^{\infty} k_p dk_p \kappa\left[ \dfrac{i\omega}{a}, \dfrac{k_p e^{a\xi}}{a}\right] \kappa\left[ \dfrac{i\omega}{a}, \dfrac{k_p e^{a\xi}}{a}\right] = \dfrac{\pi a \omega }{2 \sinh{\dfrac{\pi \omega}{a}}}~,
   \label{PLB2}
\end{equation} 
in the above yields
\begin{equation}\label{PLB4}
    \begin{split}
        \gamma(\omega_0) & = \dfrac{2}{a(2 \pi)^3} \int_{-\infty}^{\infty} d\Delta\eta \int_0^{\infty} d\omega  e^{i \omega_0  \Delta\eta} \left( \dfrac{e^{-i \omega \Delta\eta}  e^{\dfrac{\pi \omega}{a}}+ e^{i \omega \Delta\eta}   e^{-\dfrac{\pi \omega}{a}}}{1-e^{-\beta \omega}}
         +  \dfrac{e^{i \omega \Delta\eta}  e^{\dfrac{\pi \omega}{a}}+ e^{-i \omega \Delta\eta}   e^{-\dfrac{\pi \omega}{a}}}{e^{\beta \omega}-1} \right) \dfrac{\pi a \omega }{2 \sinh{\dfrac{\pi \omega}{a}}} \\
         & = \dfrac{4 \pi}{a(2 \pi)^3} \int_0^{\infty} d\omega  \left\{ \delta(\omega-\omega_0)\left[ \dfrac{e^{\dfrac{\pi \omega}{a}}}{1-e^{-\beta \omega}} +  \dfrac{e^{\dfrac{-\pi \omega}{a}}}{e^{\beta \omega}-1} \right] + \delta(\omega+\omega_0) \left[ \dfrac{e^{\dfrac{-\pi \omega}{a}}}{1-e^{-\beta \omega}}+  \dfrac{e^{\dfrac{\pi \omega}{a}}}{e^{\beta \omega}-1} \right] \right\} \dfrac{\pi a \omega }{2 \sinh{\dfrac{\pi \omega}{a}}}~. 
    \end{split}
\end{equation}
Since both $\omega_0$ and $\omega$ are positive, the second Dirac-delta term gives vanishing contribution. So we obtain (\ref{T3}) by performing the integration over $\omega$.

%\section{Calculation of $\gamma(-\omega_0)$}\label{App2}
Similarly $\gamma(-\omega_0)$ can be calculated. In this case we have
%\begin{eqnarray}
 %         \gamma(-\omega_0) & = &\int_{-\infty}^{\infty} d\Delta\eta  e^{-i \omega_0  \Delta\eta}  G^+_{\beta R}
%          \nonumber
%          \\
 %        & = & \dfrac{2}{a} \int_{-\infty}^{\infty} d\Delta\eta \int_0^{\infty} d\omega  e^{-i \omega_0  \Delta\eta} \left( \dfrac{e^{-i \omega \Delta\eta}  e^{\dfrac{\pi \omega}{a}}+ e^{i \omega \Delta\eta}   e^{-\dfrac{\pi \omega}{a}}}{1-e^{-\beta \omega}}
  %       +  \dfrac{e^{i \omega \Delta\eta}  e^{\dfrac{\pi \omega}{a}}+ e^{-i \omega \Delta\eta}   e^{-\dfrac{\pi \omega}{a}}}{e^{\beta \omega}-1} \right) 
   %      \nonumber
    %     \\
     %   &\times& \int_0^{\infty} \dfrac{2 \pi k_p dk_p}{(2 \pi)^4} \kappa\left[ \dfrac{i\omega}{a}, \dfrac{k_p e^{a\xi}}{a}\right] \kappa\left[ \dfrac{i\omega}{a}, \dfrac{k_p e^{a\xi}}{a}\right] ~.
%\end{eqnarray}
%Next use of (\ref{PLB2}) gives
\begin{equation}
      \gamma(-\omega_0)  = \dfrac{4 \pi}{a(2 \pi)^3} \int_0^{\infty} d\omega  \left\{ \delta(\omega+\omega_0)\left[ \dfrac{e^{\dfrac{\pi \omega}{a}}}{1-e^{-\beta \omega}} +  \dfrac{e^{\dfrac{-\pi \omega}{a}}}{e^{\beta \omega}-1} \right] + \delta(\omega-\omega_0) \left[ \dfrac{e^{\dfrac{-\pi \omega}{a}}}{1-e^{-\beta \omega}}+  \dfrac{e^{\dfrac{\pi \omega}{a}}}{e^{\beta \omega}-1} \right] \right\} \dfrac{\pi a \omega }{2 \sinh{\dfrac{\pi \omega}{a}}}~,
\end{equation}
which is same as (\ref{PLB4}) with $\omega_0 \to -\omega_0$.
Here the first Dirac-delta term does not contribute and therefore we find the expression for $\gamma(-\omega_0)$ as given in (\ref{T31}).

\end{widetext}

\bibliographystyle{apsrev}
\bibliography{bibtexfile}

\end{document}